\documentclass[aps,prc,twocolumn]{revtex4}

\begin{document}

\title{Intelligent Design in the Physics Classroom?}
\author{Travis Norsen}
\affiliation{Marlboro College \\ Marlboro, VT  05344 \\ norsen@marlboro.edu}

\date{\today}

\begin{abstract}
This fun polemical piece was written several months ago on a tip that
the \emph{Chronicle of Higher Education} might be interested in
publishing something like it.  Sadly (both for me and, I think, for
the \emph{Chronicle}'s readership) the editors didn't think it was
of sufficient interest to the wider academic community.  I am posting
it here at the arxiv so that it can, nevertheless, be publicly 
available.  If anyone out there wants to (suggest a place to) publish
the piece, I'm all ears.
\end{abstract}

\maketitle

A dangerous enemy has infiltrated our science classrooms and is 
infecting our students' minds.  The enemy is a profoundly unscientific 
theory masquerading as legitimate science.  Its presence in the
science classroom blurs the distinction between real science and 
arbitrary dogma and ``makes students stupid'' by leaving them less 
able to distinguish reasonable ideas from unreasonable ones -- a 
skill that is surely one of the main goals of teaching science in 
the first place.

You probably suspect the enemy I'm talking about is Intelligent Design 
(ID).  Yes, ID has infiltrated some science classrooms.  Yes, ID is 
specifically designed to blur the distinction between real science and 
religious dogma.  And yes, the phrase ``makes students stupid'' is
straight 
out of Pennsylvania Judge John Jones' recent finding that ``ID is not 
science'' and shouldn't be taught in the biology classroom.  But, in 
part because of Jones' excellent analysis, I don't think ID is a 
terribly significant danger.  It is too transparently unscientific, 
too widely recognized for what it really is: 
a thinly-veiled attempt to inject religious 
creationism into the science classroom.  

The enemy I'm worried about is something else -- something just as 
unscientific as ID, but more dangerous because it is not widely 
recognized as such: the Copenhagen interpretation of quantum mechanics.

The Copenhagen interpretation, so named because of the Danish roots 
of its main author Niels Bohr, grew out of the paradoxical nature 
of sub-atomic particles revealed by experiments in the 1920's: 
electrons sometimes acted like particles but sometimes like waves.  
This is a paradox because particles are, by definition, localized 
entities that follow definite trajectories while waves are not 
confined to any particular path or region of space.  How could the 
same thing be both confined and not confined, both a particle and a 
wave?  Paradox indeed!

Luckily, the two conflicting aspects never appear simultaneously.  
The experimental situations in which the particle and wave properties 
manifest themselves are, in a sense, mutually exclusive.  The famous 
Heisenberg Uncertainty Principle codifies this separation:  any 
experiment which reveals the precise particle character of an 
electron will necessarily obscure the wave character completely, 
and vice versa.  

If one wants to achieve a coherent, physical understanding of the 
nature of the electron, however, this is not very satisfying.  
Bohr's approach was not so much to resolve the paradox as to embrace 
it.  Naming his philosophy ``complementarity,'' he posited that the 
electron's wave 
and particle natures were mutually incompatible -- yet still jointly 
exhaustive -- perspectives.  A complete theoretical description of 
the electron would have to include both wave and particle aspects; 
yet, like the experimental situations in which they are revealed, 
the very concepts of ``wave'' and ``particle'' could not be applied 
simultaneously.  According to the Copenhagen view, physicists can 
never really understand the surprising experimental results or the 
real nature of the electron.  We must simply embrace the paradox 
and quit hoping for a coherent physical picture.

This all probably sounds rather weird and philosophical, at least 
compared to what scientists normally consider scientific.  One might 
think, therefore, that Bohr's ideas could have had little or no 
impact on the actual scientific theory of quantum mechanics.  This, 
however, is definitely not the case.  Bohr's ideas were tremendously 
influential in the development of the theory, and continue to be 
taught -- in all the textbooks and in the overwhelming majority of 
classrooms -- as an essential, ineliminable part of the formal 
textbook theory.

Indeed, Bohr's paradox-embracing philosophy has an exact counterpart 
in the theory's mathematics.  It describes electrons as waves that 
obey Schr\"odinger's wave equation.  So far so good.  But this part 
of the dynamics only applies when nobody is looking.  When somebody 
looks (i.e., when a ``measurement'' of the electron is made) it suddenly 
(one is tempted to say, magically) becomes a particle -- a process
governed, 
not by Schr\"odinger's equation, but by a different, incompatible bit of 
mathematics.  According to the Copenhagen theory, the fundamental laws 
of nature governing electrons are deeply dependent on the
human-centered concept of ``measurement.''

Bohr's colleague Pascual Jordan expressed the implications of the 
Copenhagen theory this way:  ``we ourselves produce the results of 
measurement...  We compel [the electron] to assume a definite
position; 
previously it was, in general, neither here nor there; it had not yet 
made its decision for a definite position.''

Heisenberg explains that ``we can no longer speak of the behavior of
the particle independently of the process of observation.  As a final 
consequence, the natural laws formulated mathematically in quantum 
theory no longer deal with the elementary particles themselves but 
with our knowledge of them.  Nor is it any longer possible to ask 
whether or not these particles exist in space and time objectively.''
He concludes that ``science no longer confronts nature as an objective 
observer, but sees itself as an actor in this interplay between man 
and nature.''
  
Bohr advocates complete surrender:  ``There is no quantum world... It
is wrong to think that the task of physics is to find out how Nature
is.  Physics concerns what we can say about Nature.''

I think that on some level, most physicists recognize the irrational 
and unscientific character of these sorts of statements -- but also 
that they are reasonable extrapolations from the Copenhagen theory.  
This is probably why physicists have developed a kind of pragmatic, 
anti-philosophical attitude, and why they deliberately suppress 
discussion of the more philosophical aspects of the Copenhagen 
quantum theory.  This attitude is best expressed in the popular 
slogan ``Shut up and calculate,'' often wielded against students 
wishing to steer discussion toward these interesting (if disturbing) 
implications.

But if the textbook theory really has these crazy implications, it 
is rather pathetic to just ignore and suppress them.  Unscientific 
views should be openly identified, challenged, and rejected.  Why 
haven't physicists been willing to do this?

Because they think there is no better alternative.  As physicist 
Murray Gell-Mann said, ``Niels Bohr brainwashed a whole generation of 
physicists into believing that the problem [of interpreting quantum 
theory] was solved fifty years ago.''  Basically, the orthodox dogma 
is that the Copenhagen approach is the only way to deal with the 
paradoxes. Physicists were allegedly forced -- by incontrovertible 
experimental data -- to accept Bohr's interpretation.  This is the 
premise behind physicists' pathetic and evasive strategies for dealing 
with the Copenhagen theory and its implications. 

But, in fact, this premise is a complete fabrication.  
The Copenhagen philosophy is \emph{not} the only possible conceptual 
framework for quantum theory.  There exists a completely normal, 
scientific, common-sensical alternative -- a theory that agrees 
with all of the experiments but avoids completely the unscientific 
philosophical baggage and subjectivist implications of the 
Copenhagen approach.  This alternative theory gives no special 
dynamical role to ``measurement,'' in no way implies that the 
world doesn't exist until somebody looks at it, and completely 
undermines the case for mind-over-matter anti-realism, channeling, 
the magical healing power of crystals, and all the other nonsense 
(as expressed, for example, in the bizarre recent movie \emph{What the 
Bleep do We Know?}) that draws its lifeblood from the Copenhagen 
philosophy.

This alternative theory was first proposed in the 1920's by Louis 
de Broglie, who (tragically) abandoned his ideas in the face of 
tremendous peer pressure from the likes of Bohr and Heisenberg.  
De Broglie's theory was then independently rediscovered in 1952 
by David Bohm, and clarified and elaborated in the 60's and 70's 
by John Bell.  

How does this alternative theory resolve the basic wave-particle 
paradox which spawned such bizarre contortions in the Copenhagen 
approach?  The solution is almost embarrassingly simple.  Bell explains:  
``While the founding fathers agonized over the question
\begin{center}`particle' or `wave'\end{center}
de Broglie in 1925 proposed the obvious answer
\begin{center}`particle' and `wave.' '' \end{center}

And that's that.  The paradox is resolved:  there are two entities, 
a wave and a particle.  The motion of the particle is affected by the 
wave according to a simple dynamical equation, and the resulting 
particle trajectories are completely consistent with what is observed 
in experiments.  It is hard not to agree with Bell's judgment that 
``this idea seems to me so natural and simple, to resolve the 
wave-particle dilemma in such a clear and ordinary way, that 
it is a great mystery to me that it was so completely ignored.''

And it continues to be ignored.  The theory is rarely mentioned in 
textbooks -- or , when mentioned, usually dismissed as flawed,
impossible, 
or inconsistent, all as part of a bogus proof that the Copenhagen 
view must be accepted.  But the theory exists.  It is possible; 
it is consistent; it is real.  And there is no defensible reason 
that it should not be more widely known -- i.e., more widely 
included in the quantum physics curriculum.  

This may seem like a rather technical issue that physicists should 
straighten out for themselves, an issue that non-physicist academics
shouldn't or needn't worry about.  But the academic community 
at large has a legitimate interest and stake in this issue, just as 
it has a legitimate interest and stake in the debate over Intelligent 
Design.  Like ID, Copenhagen quantum mechanics ``makes students
stupid.''  
Like ID, it probably has no place in college science classrooms.  
If it is nevertheless to be given such a place, shouldn't the 
obviously more rational alternative theory of de Broglie and Bohm 
also be taught?  
 
This is a question physicists should have asked long ago.  Given 
their stubborn refusal to do so, perhaps it is time for their 
colleagues and administrators -- and any willing Pennsylvania 
judges -- to provide the necessary wake-up call.  Because, if
you ask me,
our physics students deserve a more intelligently designed 
curriculum.

\end{document}